\documentclass[runningheads]{llncs}

\usepackage[T1]{fontenc}
\def\doi#1{\href{https://doi.org/\detokenize{#1}}{\url{https://doi.org/\detokenize{#1}}}}
\usepackage{graphicx}
\usepackage{subfigure}
\usepackage{xcolor}
\usepackage{hyperref} % for \url{}
\usepackage[english]{babel}

\newcommand{\tocheck}[1]{\textcolor{orange}{#1}}

\begin{document}
\title{Characterization of different user behaviors for demand response in data centers}
\titlerunning{User behaviors for demand response in data centers}

\author{Maël Madon%\orcidID{0000-0001-9476-4682}
\and
Georges Da Costa%\orcidID{0000-0002-3365-7709} 
\and
Jean-Marc Pierson%\orcidID{0000-0001-8948-0474}
}
\authorrunning{M. Madon et al.}

\institute{IRIT, Université de Toulouse, CNRS, Toulouse INP, UT3, Toulouse, France
\email{\{mael.madon,georges.da-costa,jean-marc.pierson\}@irit.fr}}

\maketitle

\begin{abstract}
%The abstract should briefly summarize the contents of the paper in 150--250 words.
Digital technologies are becoming ubiquitous %more and more important in our societies, as is their environmental impact.
while their impact increases. A growing part of this impact happens far away from the end users, in networks or data centers, contributing to a rebound effect. A solution for a more responsible use is therefore to involve the user. As a first step in this quest, this work considers the users of a data center and characterizes their contribution to curtail the computing load for a short period of time by solely changing their job submission behavior.  

%To this end, we use a simulation on real data from the Parallel Workloads Archive.
The contributions are: (i)~an open-source plugin for the simulator Batsim to simulate users based on real data; (ii)~the exploration of four types of user behaviors to curtail the load during a time window namely \textit{delaying}, \textit{degrading}, \textit{reconfiguring} or \textit{renouncing} to their job submissions. We study the impact of these behaviors on four different metrics: the energy consumed during and after the time window, the mean waiting time and the mean slowdown. We also characterize the conditions under which the involvement of users is the most beneficial.
% ~ 170 words last time I checked

% TODO look for keywords
\end{abstract}

\section{Introduction}
Digital technologies are increasingly contributing to global warming for instance through mining of their components, transport along their supply chains or electricity consumed during their use phase. A recent review of estimates~\cite{freitagRealClimateTransformative2021} puts this impact at 1.0–1.7 GtCO$_2$e in 2020, ie., 1.8\%-2.8\% of global greenhouse gas emissions. Authors also argue that although progress in energy efficiency of these technologies will probably continue, it will likely be outbalanced by growth in usage, leading to an overall increase of the carbon footprint. This so-called ``rebound effect'' seems difficult to fight within our research area (scheduling and distributed computing) where the focus is on energy optimization that must be effortless to end-users. On the contrary, we argue that users of digital technologies must be brought back into the loop, made aware of their impact and empowered to mitigate it.

Involving the user for environmental-aware scheduling in data centers has two aspects. One is to consider user \textit{requests} for more environment-friendly services (eg., guarantees, green labels) and try to achieve them. The other is to consider the users as a \textit{lever} for flexibility in the scheduling, ie., they accept to compromise occasionally on  their quality of service to allow some optimizations. The degradation can be spatial~\cite{guyonInvolvingUsersEnergy2019} (reducing the amount of resources allocated for the jobs), temporal~\cite{orgerieWattsYourGrid2008} (delaying their execution) or both~\cite{guyonEnergyEfficientIaaSPaaS2018}. %SMALL ,basmadjianMakingDataCenters2018}. 

This paper proposes an experimental analysis of such user levers in a context of demand response management by investigating the following question: from the users' perspective, what is the room for manoeuvre to curtail the load on the data center for a short period of time?

The rest of the article is organized as follows. Section~\ref{sec:model} presents our data center model and lists the user behaviors studied for demand response. Section~\ref{sec:exp_setup} describes the experimental setup for characterizing these behaviors. The results are presented in Section~\ref{sec:results} while Section~\ref{sec:discussion} provides a discussion on the results and the limitations of the study. Section~\ref{sec:related_works} discusses the related works. Finally, we conclude in Section~\ref{sec:conclu} and provide perspectives for future works.
% 1 page

\section{Model}
\label{sec:model}

%SMALL \paragraph{Context: data center demand response} %\mm{does that fit here? -> eventuellement moitié ici moitié Related Works}
Data centers are viewed as good candidates to participate in demand response programs~\cite{wiermanOpportunitiesChallengesData2014}. Large consumers of electricity, they also have a more flexible load than other industrial facilities. Demand response consists of adapting the electricity \textit{consumption} in response to the availability of \textit{production}. For example, some electricity markets have Coincident Peak Pricing programs where industrial consumers are charged a very high price during the time window when the most electricity is requested overall in the grid. These peak pricing events last typically 15 minutes~\cite{zarnikauResponseLargeIndustrial2013} or one hour~\cite{liuDataCenterDemand2013} but are only known \textit{afterwards}, eg., at the end of the month. The electricity supplier would only send warnings to the consumer that a peak load event may happen in the next few hours. 
%\jmp{Bizarre la phrase précédente : afterwards, mais tu dis qu'ils reçoivent après. Je comprends pas.} -> tu as raison, changé.

In our model, a demand response event will be represented by a time window of few hours (called ``demand response window'') during which the objective is to reduce electricity consumption. The event is supposed unknown in advance.
In order to characterize the efficiency of different user behaviors to react to such demand response event, we consider a data center to which users can submit their jobs. At the interface between the two is the RJMS (Resource and Job Management System), the scheduler in charge of job placement and resource management. In this section, we describe the different components of our system. %SMALL (see illustration later in Fig.~\ref{fig:simulated_system}). %\todo{decide between "data center" and "platform" and align the figure and subsections}

%\geo{Je mettrai la  partie simu ailleurs en gardant ici la moitié à propos de DC, schedule, user}
%\mm{ok, et dans la figure on garde batsim quand même du coup?}

\subsection{Data center}
%A data center is a cluster of multi-core homogeneous machines. %\geo{Je mettrai la moitié de la phrase suivante (les 16 cores) dans la partie expé}. 
%Each machine can be switched on or off by the scheduler. 
%SMALL \paragraph{Energy model}

In the data center, we only take into account the energy consumption of the multi-core homogeneous machines. The power of a machine is $P_{off}$, $P_{son}$ or $P_{soff}$ if the machine is switched off, switching on or switching off, respectively. When a machine is switched on, its power is equal to $P_{idle} + N * P_{core}$ with $P_{idle}$ the power drawn by an idle machine, $N$ the number of cores in use (ie., with a job running on it) and $P_{core}$ the power drawn by each core.

%SMALL \paragraph{Job model}
A job is completely defined by its \textit{submission time}, \textit{execution time} and number of requested cores that we denote by \textit{size} in the rest of this paper. The scheduler decides the starting time for the job and the machine it will be executed on. Note that the scheduler in our model only execute jobs on single machines.
%\geo{En fait le fait d'être sur une seule machine c'est pas vraiment dans ton modèle, juste dans tes expés non ? Peut être à déplacer après (sinon ici ça passe quand même sans problème)} -> oui en effet.. c'est plutôt mon scheduler qui impose ça. Modifié un peu la phrase.
We suppose perfect communication without latency.

\subsection{Scheduler}
The scheduler is a bin-packing scheduler with shutdown (same as Guyon et al.~\cite{guyonEnergyEfficientIaaSPaaS2018,guyonInvolvingUsersEnergy2019}). It is a greedy algorithm trying to schedule (``pack'') all the jobs in the least possible machines and shut down idle machines. To do so, it maintains and updates two data structures: a queue of waiting jobs and a list of switched-on machines. The queue of jobs is sorted by \textit{decreasing} size order -- and by increasing submission time (first come first served) in case of a tie. The list of machines is sorted by \textit{increasing} order of available cores. Every time one (or more) job is submitted or finishes, the scheduler goes through the job queue in order and tries to find for each job the smallest machine where it fits. If no machine is found that way, a new machine (if any) is powered on and the job is scheduled on this machine. After that, we immediately shut down all idle machines.

\subsection{Users}
%The jobs are submitted by the users of the infrastructure.
%\geo{début phrase suivante dans expé} In the experiments, we replay a real public workload trace containing the information about the submitting user for each recorded job. 
During the demand response  window, users are asked to make an effort to curtail the load in the data center. They do so by adopting different behaviors: %SMALL described below.%SMALL  and illustrated in Fig.~\ref{fig:behaviors}.

%SMALL
%\begin{figure}
%    \centering
%    \includegraphics[width=0.8\textwidth]{figures/behaviors}
%    \caption{The five user behaviors studied}
%    % source: https://drive.google.com/file/d/1dAzotwZeRBqA4GLXQhPMPkywgfEr5UaG/view?usp=sharing
%    \label{fig:behaviors}
%\end{figure}

\begin{itemize}
    \item \textbf{rigid}: replay jobs as in the original workload; Baseline for comparison.
    \item \textbf{renounce}: do not submit jobs originally submitted during the window.
    \item \textbf{delay}: delay all job submissions to the end of the  window.
    \item \textbf{degrad}: divide the size of the jobs by two, rounded up. The execution time stays the same. Note that the rounding implies that when only one core is requested for a job, the job remains unchanged. 
    \item \textbf{reconfig}: also divide the size by two, rounded up, but increase the execution time to match the original computing mass. We make the hypothesis of perfect speedup, ie., a job executing on one core completes in exactly twice the time than on two cores.
\end{itemize}

\section{Experimental setup}
\label{sec:exp_setup}

%SMALL This section describes the software, workload, platform and experimental choices used for the experiments. 

%SMALL
%\begin{figure}
%    \centering
%    \includegraphics[width=.8\textwidth]{figures/simulated_system}
%    \caption{The simulated system}
%    \label{fig:simulated_system}
%    % source: https://drive.google.com/file/d/1dAzotwZeRBqA4GLXQhPMPkywgfEr5UaG/view?usp=sharing
%\end{figure}

\subsection{Software used for simulation}
To simulate our system, we use Batsim~\cite{dutotBatsimRealisticLanguageIndependent2016}, an open-source infrastructure and resource management system simulator\footnote{Batsim: \url{https://batsim.org/}} based on SimGrid\footnote{SimGrid: \url{https://simgrid.org} with the energy plugin \url{https://simgrid.org/doc/latest/Plugins.html?highlight=energy\#host-energy}}.
%SMALL. For the data center simulation (including energy consumption), Batsim relies on SimGrid\footnote{SimGrid: \url{https://simgrid.org}}.%SMALL, widely used and trusted in the community.
We implemented the bin-packing scheduler for Batsim in C++. We also developed a plugin called ``batmen'' to interact with simulated users and receive their job submissions. For the purpose of this study, users replay an input workload trace except in the demand response window where they act according to their behavior. We therefore implemented five user classes corresponding to the five behaviors. %SMALL of Fig.~\ref{fig:behaviors}.
Our code is open source\footnote{code repository: \url{https://gitlab.irit.fr/sepia-pub/mael/batmen}}. 
With this simulation tool, we designed and conducted an experimental campaign whose main details are given below\footnote{All scripts are available to reproduce our results: \url{https://gitlab.irit.fr/sepia-pub/open-science/demand-response-user}}.%SMALL For the sake of reproducibility, a repository containing the software environment, the scripts used to launch the experiments and to analyse the output data is also provided\footnote{experiment repository: \url{https://gitlab.irit.fr/sepia-pub/open-science/demand-response-user}}. % todo: change the URL to europar2022 if article accepted
%SMALL The reader will find more specific details there.
%\jmp{il ne faut pas cacher les liens en blinded reviews non ?}

\subsection{Workload}
\label{subsec:workload}
% next sentence: moved here from Model
%SMALL In the experiments, w
We replay a real public workload trace containing the information about the submitting user for each recorded job. We chose the 2-year trace from MetaCentrum (national grid of the Czech republic), available in the Parallel Workload Archive\footnote{METACENTRUM-2013-3.swf available at \url{https://www.cs.huji.ac.il/labs/parallel/workload/l_metacentrum2/index.html}}. %SMALL As mentioned in the original paper releasing the log~\cite{klusacekRealLifeExperienceMajor2017}, t
The platform is very heterogeneous and underwent majors changes during the logging period\cite{klusacekRealLifeExperienceMajor2017} so
%SMALL. For the purpose of our study, 
we perform the following selection:
\begin{enumerate}
    \item We truncate the workload to keep only 6 months (June to November 2014) where no major change was performed in the infrastructure and we remove all the clusters whose nodes have more than 16 cores;
    \item From this truncated workload, we remove all jobs with an execution time greater than one day and all jobs with a size greater than 16. It leaves us with a workload manageable with machines of a usual size, and without more than one day of inertia. 
\end{enumerate}

\subsection{Platform}
%SMALLAccording to the system specifications released with the log, we counted that t
The first selection step keeps a total of 6304 cores. %We also measured that t
The second selection step exclude 2.7\% of jobs from the truncated workload, representing 73.7\% of the mass (in core-hour). Consequently, we create a simulated platform adapted to this load with $6304 * (1 - 0.737) / 16 =$ \textbf{104 homogeneous 16-core machines}.
Power constants ($P_{idle}=100W$, $P_{core}=7.3125W$, $P_{off}=9.75$, $P_{son}=100W$ and $P_{soff}=125W$) for the servers and time to switch on ($T_{son}=150s$) and switch off ($T_{son}=6s$) are measurements in Taurus Grid'5000 cluster from existing work~\cite{guyonInvolvingUsersEnergy2019}.

%SMALL Power constants are given in Table~\ref{tab:energy_cons}.%\jmp{dans la Table ci-dessous, tu n'as pas défini avant dans le modèle ce qu'est Tson et TSoff}

%\begin{table}
%    \centering
%    \begin{tabular}{|c|c|c|c|c|c|c|}
%        $P_{idle}$ & $P_{core}$ & $P_{off}$ & $P_{son}$ & $P_{soff}$ & $T_{son}$ & $T_{soff}$ \\
%        \hline 100 W & 7.3125 W & 9.75 W & 100 W & 125 W & 150 s & 6 s
%    \end{tabular}
%    \caption{Power constants for the servers and time to switch on ($T_{son}$) and switch off ($T_{son}$). Measurements in Taurus Grid'5000 cluster from previous work~\cite{guyonInvolvingUsersEnergy2019}.}
%    \label{tab:energy_cons}
%\end{table}

\subsection{Experimental campaign}
\begin{figure}
    \centering
    \includegraphics[width=\textwidth]{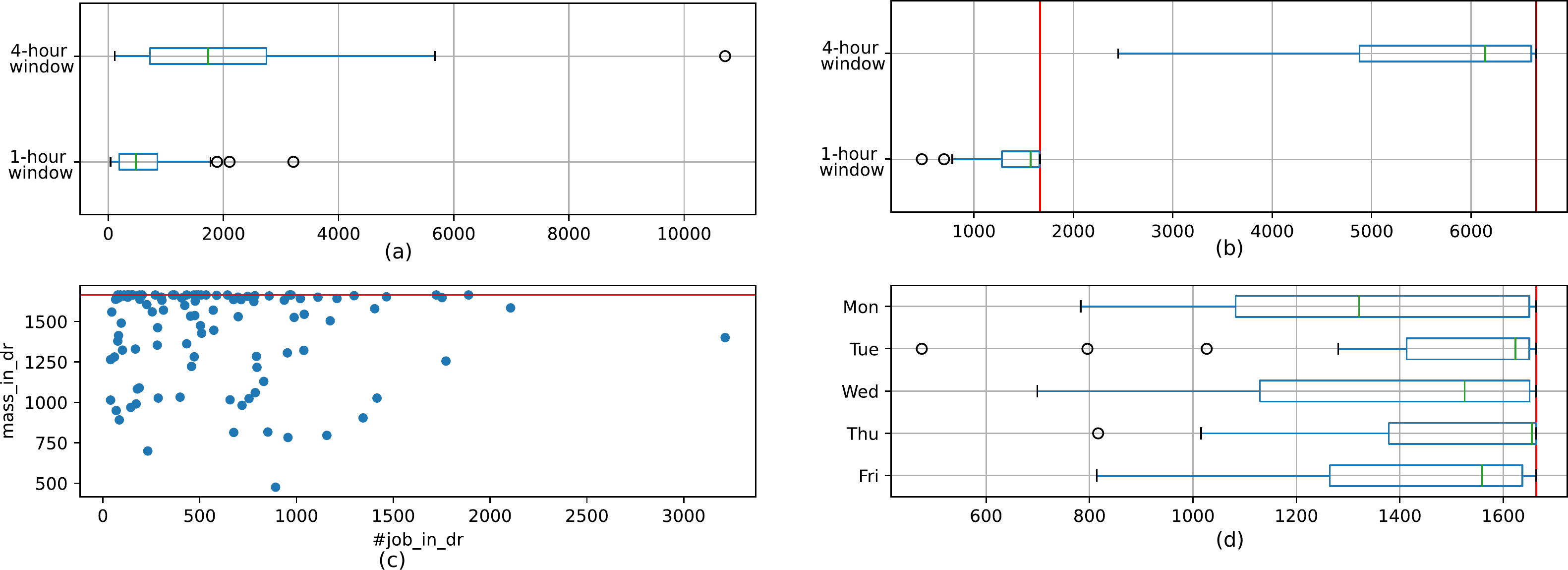}

    \caption{Descriptive statistics for the 105 experiments. Red lines corresponds to the infrastructure (1664 cores). (a)~number of jobs submitted in window; (b)~computing mass (in core-hour) in window; (c)~computing mass in window by number of submitted jobs (1-hour window); (d)~computing mass in window by weekday (1-hour window)}
    %\jmp{bizarre la légende : each dot ... alors que ce sont des boxplots, il n'y a pas de points. En plus on dirait que les figures se marchent dessus}
    \label{fig:data_descriptive_stats}
\end{figure}

We conducted an experimental campaign consisting of 105 experiments (the number of weekdays between Jun 1, 2014 and Oct 23, 2014). For each experiment, we vary the input workload corresponding to three full days of data center operation. We make the demand response event arise at 16:00 on day 2, chosen to be a weekday. This choice is justified by a characterization of 26 years' coincident peak pricing data~\cite{liuDataCenterDemand2013}, given that the MetaCentrum trace also displays diurnal and weekday/weekend patterns. We study two lengths for the demand response window: one and four hours. We also tried other starting times (drawn at random) and other window lengths (0.5 and 2 hours) but decided not to report their results here as they are not leading to different conclusions. %SMALL Note that the scripts remain available in the experiment repository.

The simulation starts one day before the event and stops one day after, to ensure the infrastructure runs at nominal load on day 2 and has absorbed the event by the end of day 3 (the selected jobs in the workload have an execution time lower than one day). In each experiment we simulate the five user behaviors on the two windows with the user behavior ``rigid'' as a baseline. Descriptive statistics on the experiments %SMALL(number of jobs submitted in the window, load of the infrastructure for the baseline behavior) 
are displayed in Fig.~\ref{fig:data_descriptive_stats}. 

%SMALL The advantage of doing simulations with an optimized tools like Batsim is the fast execution time: our
The campaign launched in parallel on a 2 x 8-core Intel Xeon E5-2630 v3 machine finished in less than two hours, based in France, and ran 2 times in total, %\jmp{utile le "and ran 2 times"?} -> c'est le paragraph type retourné par l'outil "GreenAlgorithm". Ca prend en compte le nombre de fois que t'as relancé tes XP
this has a carbon footprint of around 50 g CO2e (calculated using green-algorithms.org v2.1~\cite{lannelongueGreenAlgorithmsQuantifying2021}). 
%An other advantage is that we can rely on Batsim to give detailed outputs from which we extract the scheduling metrics while the underlying energy plugin for Simgrid\footnote{\url{https://simgrid.org/doc/latest/Plugins.html?highlight=energy\#host-energy}}  simulates energy consumption in the infrastructure.  

\section{Results}
\label{sec:results}
%SMALL The experimental campaign produces several gigabytes of output data (energy consumption, scheduling details). We summarize them in this section by looking first at the impact of the different behaviors on the energy consumption, and then on usual scheduling metrics.

\subsection{Energy metrics}

\begin{figure}
    \centering
    \includegraphics[width=.85\textwidth]{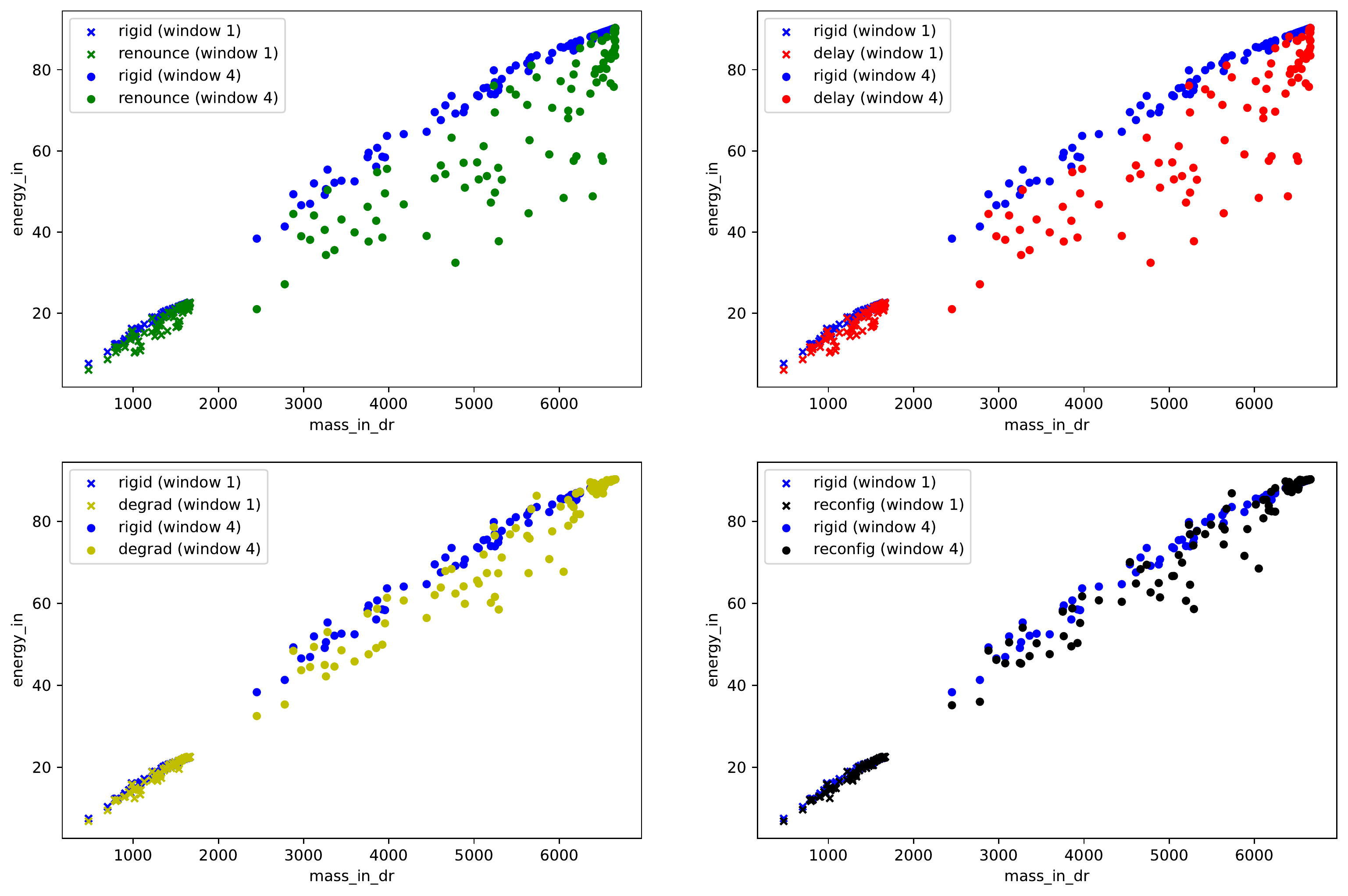}
    \caption{Energy consumed in each simulation. Y-axis: energy consumed (in kWh) during the demand response window. X-axis: computing mass (in core-hour) during the demand response window for the baseline behavior.}
    \label{fig:energy_in}
\end{figure}

We recall our research question: by intervening only on the user's side, what energy gains can be expected by adapting one's behavior for a few hours?
%\jmp{en fait, la question n'est jamais posée comme ça dans l'introduction}
Fig.~\ref{fig:energy_in} displays the \textbf{energy consumed during the demand response window} for every experiment and every behavior. Values are scattered by the total load of the infrastructure during the window for the baseline behavior. %SMALLFor that behavior, 
We note an almost linear relationship between infrastructure load and consumed energy. Deviations from the linear line are due to situations favoring a more or less good packing from the scheduler inside the 16-core machines. 
%SMALL A first observation is that the 
``renounce'' and ``delay'' behaviors perform identically for this metric: Users of both behaviors stop submitting inside the demand response window, resulting in a lower energy consumption compared to the baseline. This gain is the best we can expect. Behaviors ``degrad'' and ``reconfig'' display similar results. In addition, one would expect a positive correlation between the load of the platform and the relative energy gains of the four behaviors compared to the baseline. It would translate into an increasing distance between the colored dots and the blue dots in the graphs, as the load increases.  Counter-intuitively, this does not seem to happen.

\begin{figure}
    \centering
    \subfigure[1-hour demand response window]{
        \includegraphics[width=.8\textwidth]{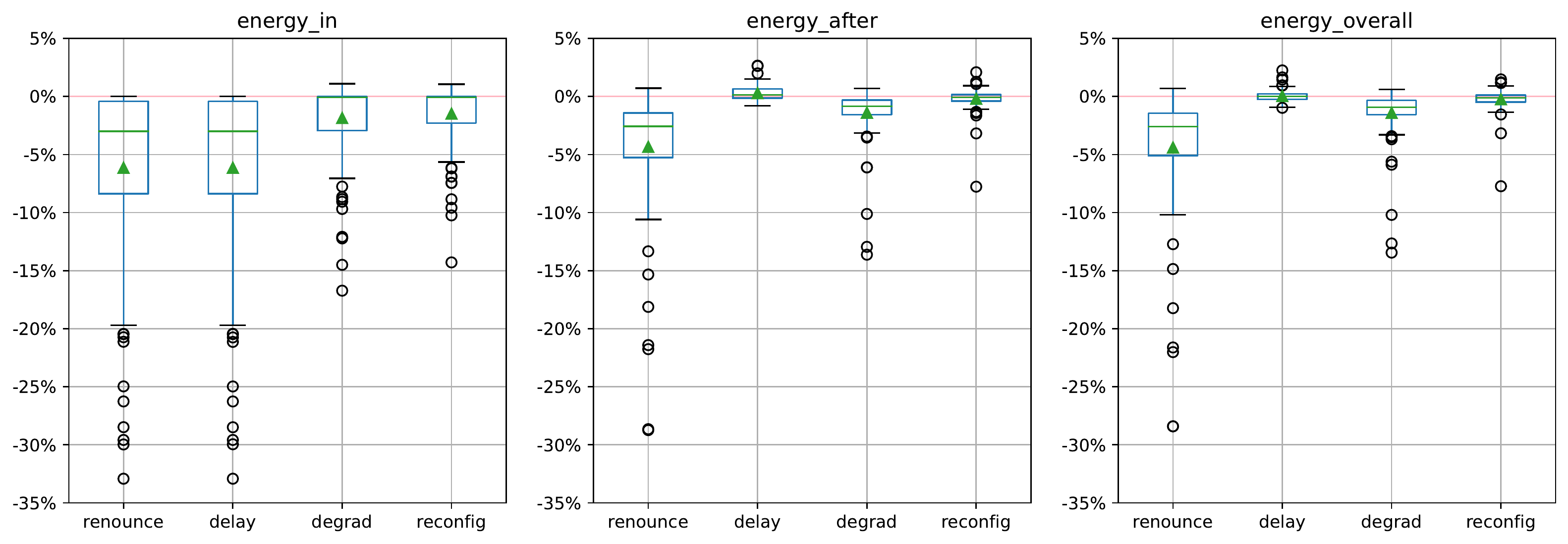}}
    \subfigure[4-hour demand response window]{
        \includegraphics[width=.8\textwidth]{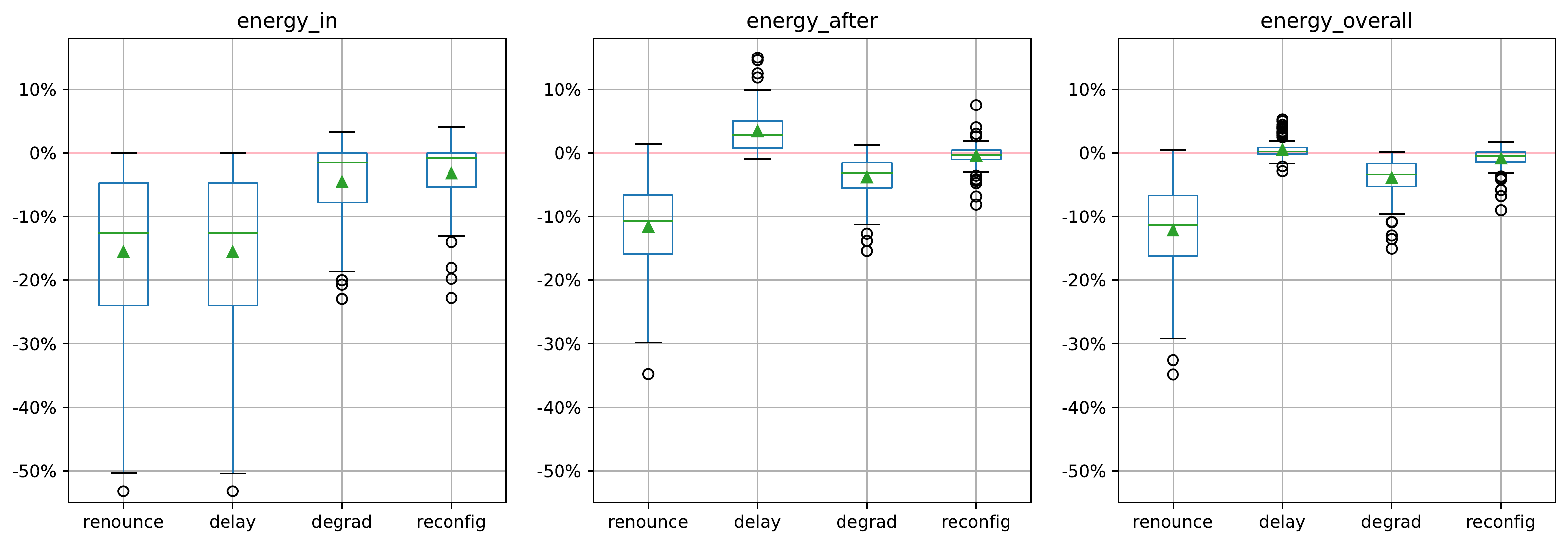}}
    \caption{Energy metrics per behavior relatively to the baseline behavior. The green triangle in the box plots indicates the mean.}
    \label{fig:energy_metrics}
\end{figure}

The experimental campaign showing very scattered results, Fig.~\ref{fig:energy_metrics} displays the relative energy gains for each experiment as box plots. We can read for example that ``renounce'', the most radical behavior, allows energy savings of up to 33\% in the window for a one-hour window, and 53\% for a four-hour window. The savings do not go up to 100\% because jobs that were already there before the window are still running in the infrastructure, which consumes energy.

In addition to the energy consumed \textit{within} the window, Fig.~\ref{fig:energy_metrics} shows the impact of the different behaviors on the energy consumed \textit{after} the demand response event ie., from 17:00 or 20:00 on day2 (depending on the window length) to 24:00 on day3. For this second metric, ``delay'' performs very differently compared to ``renounce''. All the jobs within the window get postponed, resulting in an extra power consumption at the end of the window: +0.3\% (resp. +3.4\%) on average for a 1-hour (resp. 4-hour) window. This behavior remains neutral with respect to overall energy consumption (\textit{within} + \textit{after} the window).% Interestingly, 
The behavior ``reconfig'', which also keeps a constant mass of jobs compared to the baseline, allows some optimizations. Up to 10\% overall energy consumption could be saved because the reconfigured jobs ``fit better in the holes'' left by the available cores in the switched on machines. ``Degrad'' performs unsurprisingly better in all respects, the users having accepted to reduce the mass of job submitted.

Finally, we notice that \textbf{the bigger the window, the better the energy gains}. This is due to inertia of the system: with a longer window, a behavior on the submitted jobs has more time to make a difference compared to the residual jobs that are still running in the infrastructure. 
%SMALL This effect will be analysed deeper in the Discussion.

\subsection{Perceived impact on the scheduling}
\begin{figure}
    \centering
    \subfigure[Mean waiting time (in seconds)]{
        \includegraphics[width=.9\textwidth]{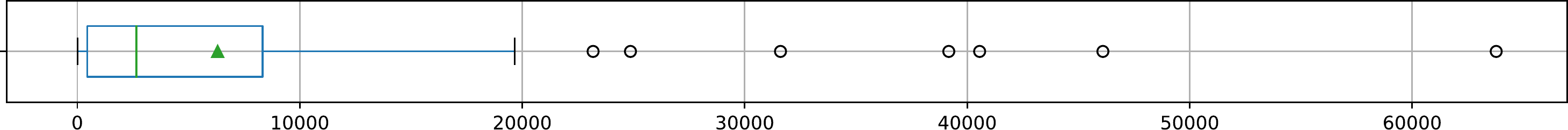}}
    \subfigure[Mean slowdown (dimensionless)]{
        \includegraphics[width=.9\textwidth]{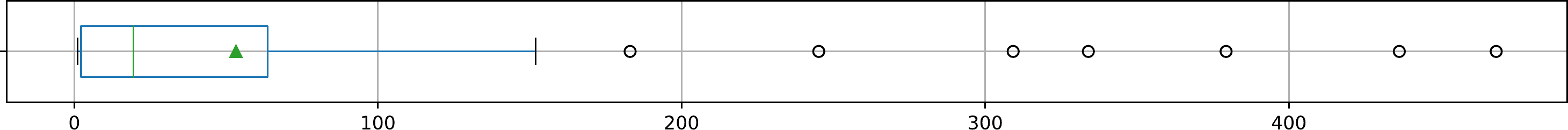}}
    \caption{Scheduling metric distribution for the 105 experiments, baseline behavior.}
    \label{fig:sched_metrics_baseline}
\end{figure}

%To study the perceived impact on the scheduling, 
We use two usual metrics: mean waiting time and mean slowdown. Waiting time is the time a user has to wait until her job starts running: $waiting_{time} = starting_{time} - submission_{time}$. Slowdown expresses this extended completion time as a function of execution time: $slowdown = (finish_{time} - submission_{time}) / execution_{time}$. For each experiment, we take the average waiting time (resp. slowdown) on all jobs submitted between the beginning of the demand response window and the end of the experiment (same period as metric \textit{energy\_in + energy\_after}). Fig.~\ref{fig:sched_metrics_baseline} shows these results %SMALL as box plots 
for the ``rigid'' behavior. We observe that for half of the experiments, the mean waiting time is below one hour (3600s) and the mean slowdown below 25. These are experiments with an unsaturated infrastructure and a queue of waiting job often empty. On the other hand, there are also cases of high congestion (eg., the seven outliers at more than 6 hours mean waiting time).

\begin{figure}
    \centering
    \subfigure[1-hour demand response window]{
        \includegraphics[width=.9\textwidth]{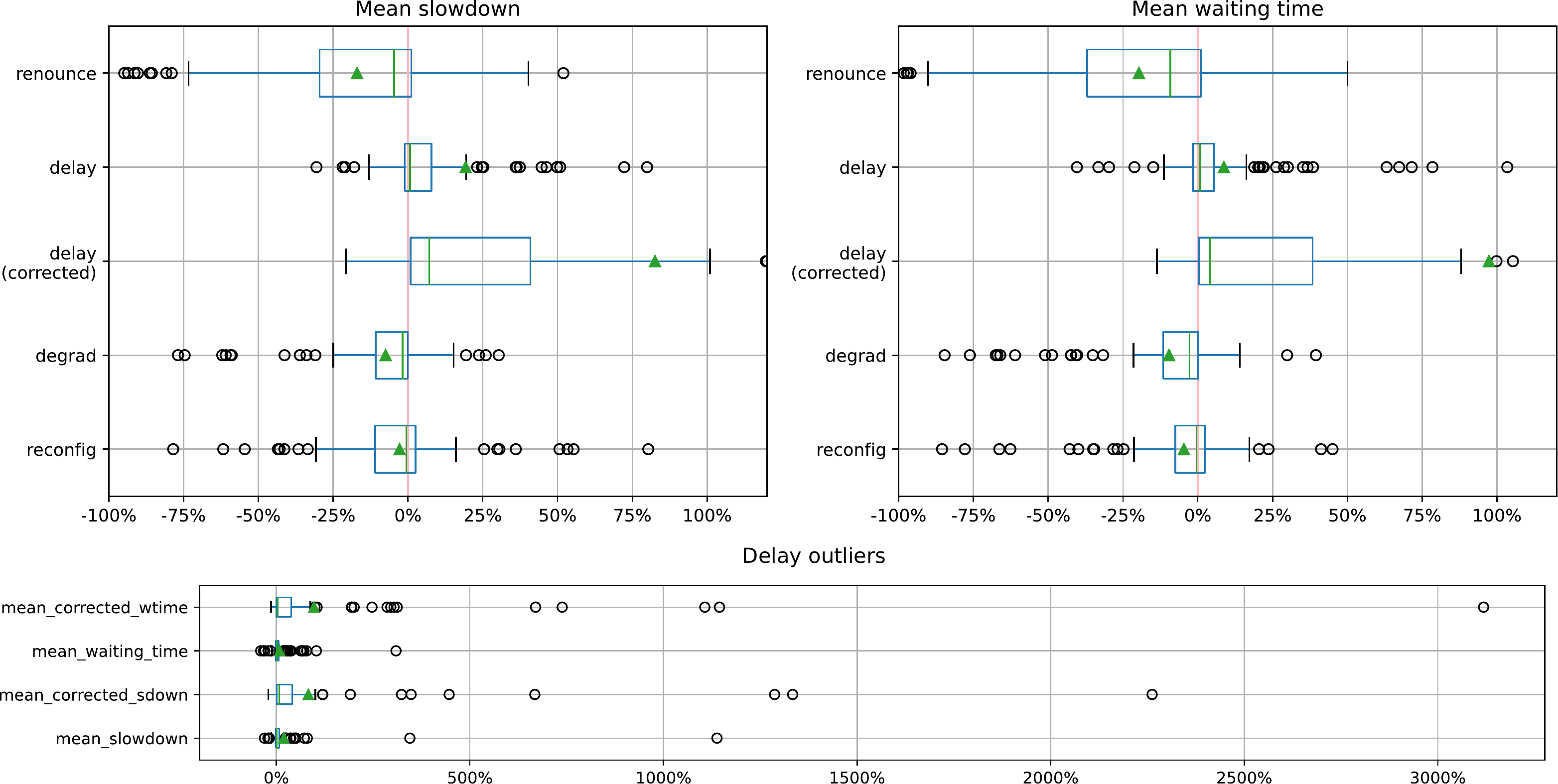}}
    \subfigure[4-hour demand response window]{
        \includegraphics[width=.9\textwidth]{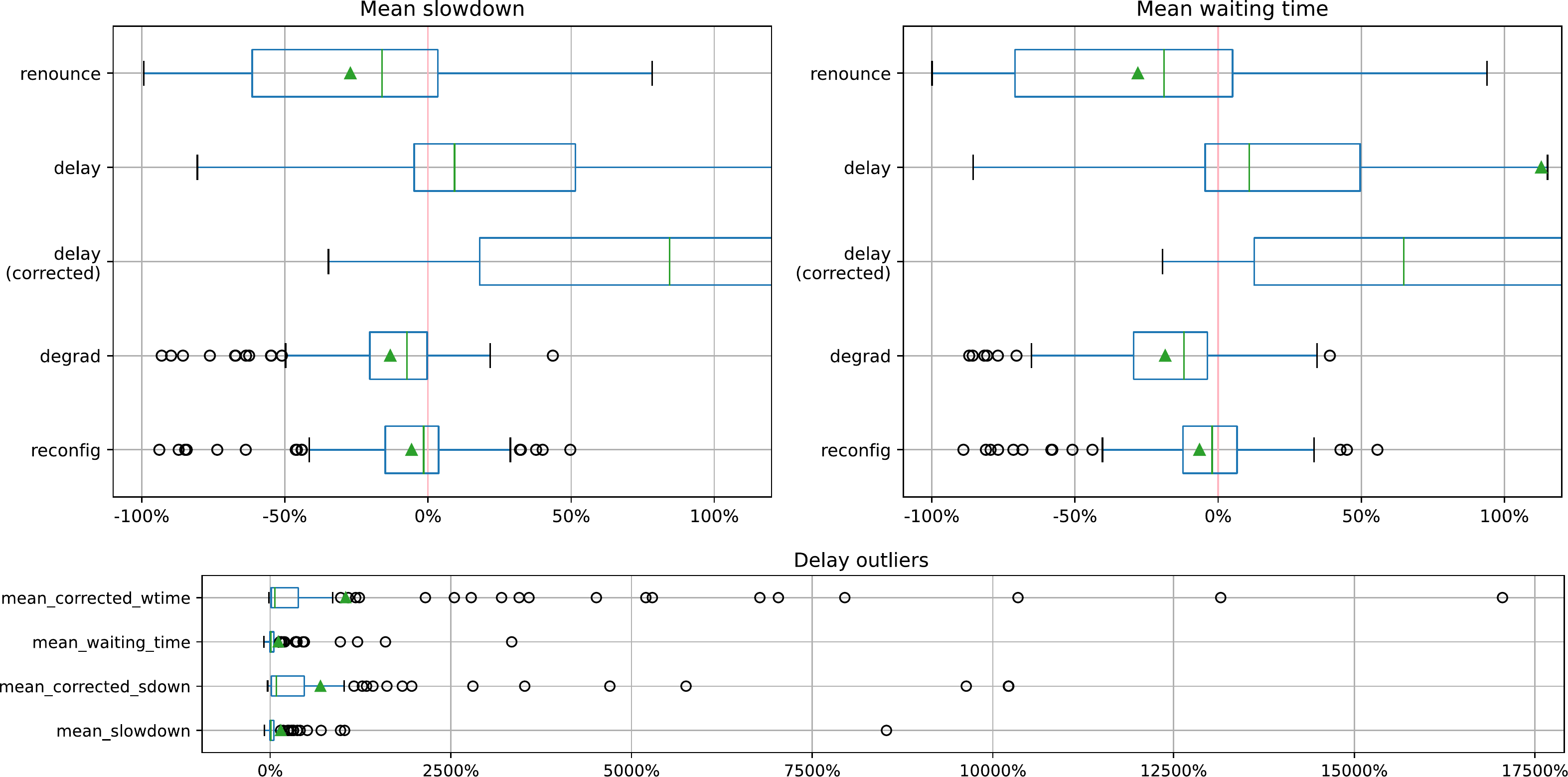}}
    \caption{Scheduling metrics per behavior relatively to the baseline behavior}
    \label{fig:sched_metrics}
\end{figure}

The results for the other behaviors are plotted in Fig.~\ref{fig:sched_metrics}, as a percentage of gain/loss compared to the baseline. Specifically for the behavior ``delay'', we provide both \textit{corrected} and \textit{uncorrected} metrics. The uncorrected slowdown and waiting time are calculated in relation to the \textit{new} (delayed) submission times while the corrected ones use the \textit{original} submission times (from the baseline). Note also that for the behavior ``renounce'' some jobs have been canceled, thus the mean waiting time and slowdown is calculated on a subset of the jobs compared to the other behaviors. From Fig.~\ref{fig:sched_metrics} it can be observed that the behaviors ``renounce'', ``degrad'' and ``reconfig'' (in this order) affects the scheduling positively on average. This is not surprising as the first two behaviors reduce the total mass of jobs to compute and the third allows a better packing. Yet, the scheduling gets worsened in a significant number of cases (around 50\% for ``reconfig'' and 25\% for ``degrad'' and ``renounce''), due to bad choices of the scheduler. 

The behavior ``delay'' stands out from the others as it affects the scheduling negatively in most cases, even for the uncorrected metrics. It gets even worse when including the extra waiting time from the delayed job in the calculation of the corrected metrics. In fact, it is preferable in terms of waiting time and slowdown that the job submissions are spread out throughout the time.

\section{Discussion}
\label{sec:discussion}
%In this section, we first provide an explanation of the results thanks to a newly introduced quantity, before summarizing the characteristics of each behavior. Then, we discuss the limitations of our study.

\subsection{The fluid-residual ratio: an explanation of the results}
\begin{figure}
    \centering % Expe18
    \includegraphics[width=.7\textwidth]{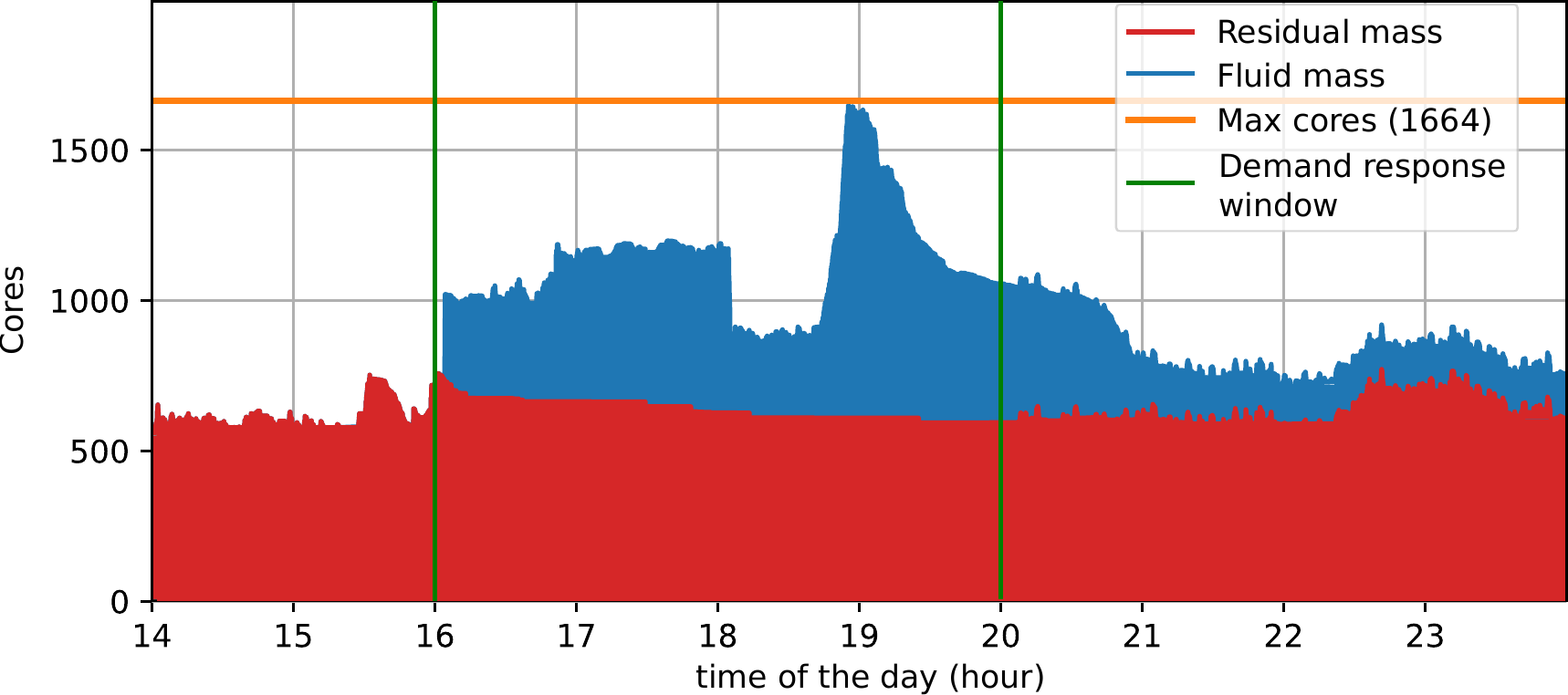}
    \caption{Example of fluid and residual mass. \textit{(Thursday Jun 26 2014)}}
    \label{fig:fluid_residual_definition}
\end{figure}

As seen previously in Fig.~\ref{fig:energy_in}, the achievable energy savings in the demand response window cannot be explained by the infrastructure load during that window. In fact, it is possible that the load is very high because of a large mass of job submitted \textit{before} the window, although the load on which the users have an influence is the mass submitted \textit{during} the window. We call these two quantities the \textbf{residual mass}, submitted outside of the window, and the \textbf{fluid mass}, submitted inside the window (Fig.~\ref{fig:fluid_residual_definition}).

\begin{figure}
    \centering
    \includegraphics[width=.8\textwidth]{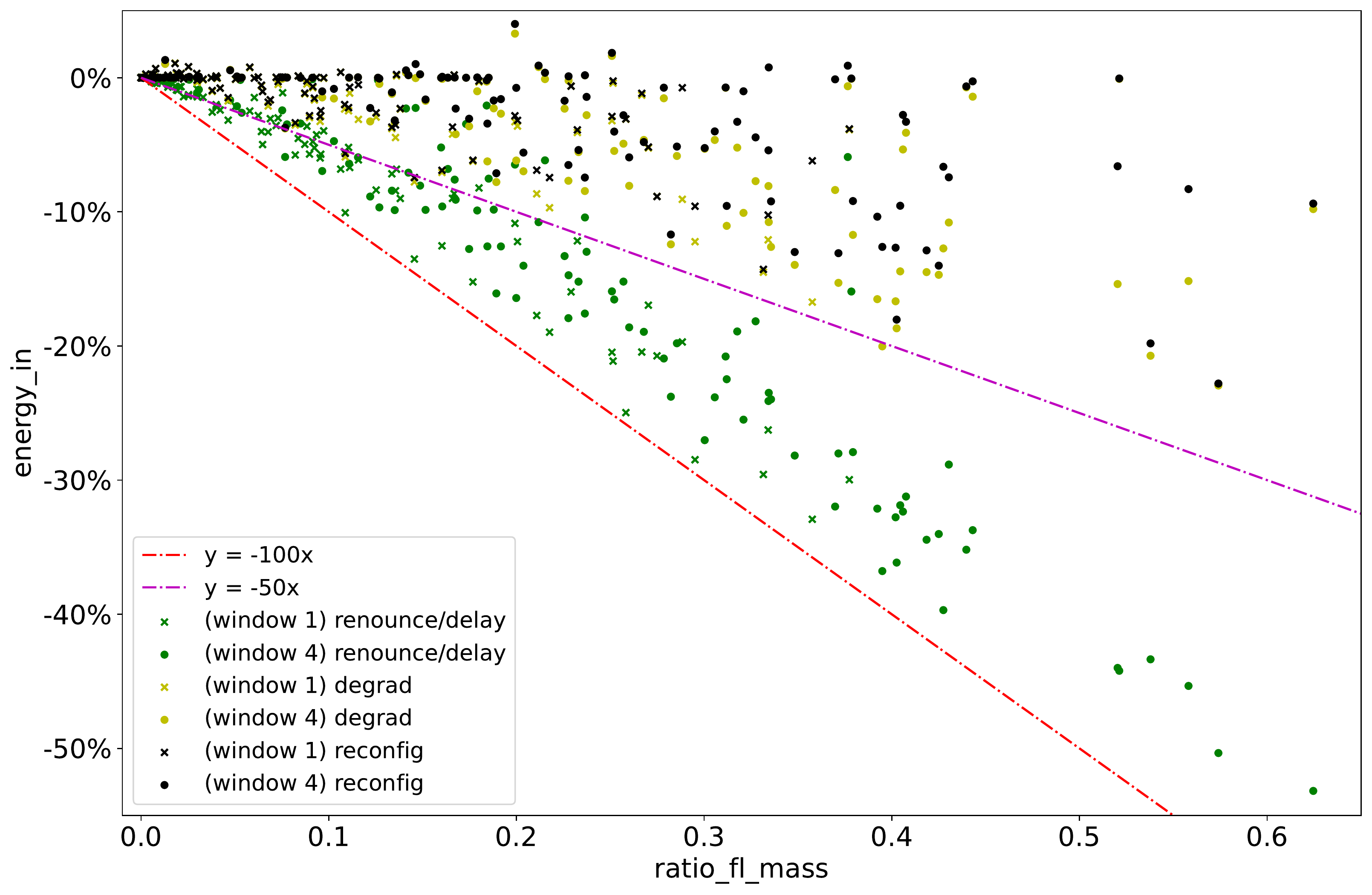}
    \caption{Energy gains in function of the fluid-residual ratio. Only one plot for the behaviors ``renounce'' and ``delay'' because they are identical for this metric.}
    \label{fig:fluid_residual_explain}
\end{figure}

Users, by accepting to ``renounce to'' or ``delay'' their jobs allow to cut the energy consumption due to the fluid mass, which is roughly proportional to the mass itself, as we saw before. In other terms, \textbf{the gains during the window are at most equal to the proportion of fluid mass in that window}. This is exactly what we see in Fig.~\ref{fig:fluid_residual_explain} displaying the energy gains as a function of the fluid-residual ratio. The red line indicates the best possible gains, which are almost achieved by ``renounce'' and ``delay'' behavior (the non-linearity of the energy model explaining the gap). 

In some cases however, these behaviors don't realize that gain: In the \textbf{saturation} cases, %when the infrastructure is at full load with 
when many jobs are waiting in the queue. The removal of the fluid mass is compensated by the execution of the awaiting residual mass.

For the ``degrad'' behavior, gains are expected to be of half the fluid mass at most, as users divide their submitted mass by two. In practice, the results are even more scattered and further away from their optimal (magenta line). This is partially due to the saturation effect, but also to rounding (eg., a job with an original size of 3 will be submitted with size 2 and a job with size 1 remains unchanged) and imperfect packing. The analysis for the behavior ``reconfig'' is similar, with even less expected gains. Some experiments even make negative gains: they are due to the greedy and non-clairvoyant scheduler taking bad decisions for the future, like switching off a machine just before the submission of new jobs.

\subsection{Pros and cons of each behavior}
%SMALL Building upon what we learnt from the results, here is a summary discussion on the characteristics of each behavior (see Table~\ref{tab:behavior_ranking}). To begin with, t
The behavior \textbf{renounce} performs the best for all the metrics studied (summary in Table~\ref{tab:behavior_ranking}). We saw that it actually reaches the optimal energy gains during the window for unsaturated cases. This rank is not surprising considering the sacrifice required from the user. Yet, we think that such a behavior is often overlooked in similar studies and argue that environmentally aware users or users provided with a proper incentive would do it. Moreover, some jobs running in data centers today might not be indispensable. 

On the other end of the spectrum, the behavior \textbf{reconfig} seems to be the most acceptable to the users as it does not decrease the mass initially submitted and provides better waiting time and slowdown than ``delay'' for both the jobs within and after the window. ``Reconfig'' is a good trade-off to achieve some optimizations with a low effort from the user, especially in combination with bin-packing schedulers and on/off policies (see~\cite{guyonInvolvingUsersEnergy2019}).

\textbf{Delay} also keeps the mass constant, which ranks it second behind ``reconfig'' in terms of acceptability. Same as ``renounce'', it reaches the optimal energy gains during the window. However, it introduces an overhead in overall energy consumption and slowdown compared to the baseline behavior. Note that this overhead would probably be less important in real life due to users adapting their behavior if they experience congestion in the infrastructure. This is the limit of blindly replaying past workload traces in simulations, as pointed by Feitelson~\cite{feitelsonResamplingFeedbackNew2016}.

Finally, the behavior \textbf{degrad} ranks second or third in all the categories of Table~\ref{tab:behavior_ranking}. It remains an interesting trade-off between simply renouncing to a job and reconfiguring it at constant mass. We can think about optional features in an application that can be cut off if needed (eg., recommendations for e-commerce, alternative paths for mapping apps).

\begin{table}[t]
    \centering
    \begin{tabular}{c|c|c|c|c}
         \textbf{behavior} & \textbf{energy in} & \textbf{energy overall} & \textbf{scheduling metrics} & \textbf{acceptability} \\ \hline
         \textbf{renounce} & 1st & 1st & 1st & 4th \\
         \textbf{delay} & 1st & 4th & 4th & 2nd \\
         \textbf{degrad} & 3rd & 2nd & 2nd & 3rd \\
         \textbf{reconfig} & 4th & 3rd & 3rd & 1st \\ 
         
    \end{tabular}
    \caption{Summary ranking of the four behaviors with regards to their impact on energy consumption and scheduling metrics. \tocheck{The column ``acceptability'' is opinion-based}, it reflects the size of the effort asked from the user. %\jmp{dans ce tableau, la notion d'acceptabilité est subjective, alors que les autres sont mesurées. Il faudrait rappeler cette subjectivité quelque part}
    }
    \label{tab:behavior_ranking}
\end{table}

All in all, having different user behaviors remains one lever for energy saving among others. It has the particularity of having some latency, which makes it not optimal in a context of demand response without prediction. \textbf{Taking into account these behaviors inside the scheduler} seems essential to make the best of their potential and go beyond the fluid-residual limit. For example in combination with malleable applications~\cite{dupontEnergyawareSchedulingMalleable2020} or contracts with the data center operator specifying the degradation the user is willing to accept~\cite{basmadjianMakingDataCenters2018}. Nevertheless, latency is not critical in other contexts, and involving the user appears as the main path towards a \textit{sufficient}~\cite{hiltyComputingEfficiencySufficiency2015} use of our technologies, if not the only.

\subsection{Limitations}
\label{sec:limits}
%The work presented in this article has some limitations that we present below in two categories: methodological and modelling limitations.

\paragraph{Model simplifications} 
In our data center simulations, we do not take into account the latency and bottleneck effects in the communications. Also, we suppose perfect speedup in the model, ie., a job executed on two cores will take exactly twice longer than the same job executed on four cores. Finally, we accounted only for the energy consumption of the CPUs, and neglected others like memory, network or cooling. Hopefully, the powerful simulation tools that we use (Batsim and Simgrid) will help us to overcome these simplifications in future works.

\paragraph{Methodological limitations}
We see three major threats to the validity of our method to answer the research question. First, we study only one scheduler (bin-packing) while results with other common schedulers (FCFS, easy-backfilling...) would have been of interest. Second, we use only one input trace (MetaCentrum) which comes from a research infrastructure and not a production cloud, and we perform a selection from it (see~\ref{subsec:workload}) that might make us miss the big picture. Finally, our study includes all the limitations related to the use of a simulation, especially when dealing with human behaviors which are unpredictable.

\section{Related works}
\label{sec:related_works}
Among the large body of work on energy-aware scheduling in data centers, some authors have studied strategies involving the users. Some works aim at providing guarantees to their users (``green offers''~\cite{gargGreenCloudFramework2011}, ``green SLA''~\cite{haqueProvidingGreenSLAs2013,amokraneGreenslaterSatisfyingGreen2015}) and commit to fulfilling them by classical methods (self-supply of renewable energy~\cite{haqueProvidingGreenSLAs2013}, geo-distributed data centers with variable PUE and energy mix~\cite{gargGreenCloudFramework2011,amokraneGreenslaterSatisfyingGreen2015}). 

More related to this paper, some works study user flexibility as a lever for energy efficiency. For example Guyon et al.~\cite{guyonInvolvingUsersEnergy2019} give to the users the choice between three execution modes (big, medium, little) for their jobs. Small execution modes request fewer resources but take longer to complete. They achieve gains through spatial consolidation with a bin-packing algorithm. Orgerie et al.~\cite{orgerieWattsYourGrid2008} save energy through thermal-aware scheduling and smart resource switch off by letting the users choose between different submission times on the basis of energy consumption estimations for each of the alternatives. A combination of both spatial and thermal consolidation is proposed in an other work by Guyon et al.~\cite{guyonEnergyEfficientIaaSPaaS2018} or in the All4Green project~\cite{basmadjianMakingDataCenters2018}, where user involvement is leveraged through contracts between the energy supplier, the data center and the user. The latter work, also in a context of demand response, is the closest to our approach. However, it integrates demand response mechanisms affecting the user with mechanisms transparent to them (use of batteries, precooling, geographical workload migration) so much so that the contribution of each user behavior to the final results is difficult to identify. 

The originality of our work is to focus on the user behaviors which allows to provide a characterization of them. To the best of our knowledge, we are also the only ones to consider the behavior of simply \textit{renouncing} to job submissions. It is a radical behavior but to be considered in a sufficiency approach.

\section{Conclusion and future works}
\label{sec:conclu}
In this paper, we study four different ways for a user of a data center to curtail her load for a certain period of time by changing submission behavior. These behaviors are delaying, degrading, reconfiguring or renouncing to the jobs during the time period. We show experimentally through simulation on real world data that these behaviors have a certain latency for decreasing the load on the infrastructure. Indeed they cannot decrease the load due to jobs that are already running on the infrastructure. Therefore, we define two quantities, the \textit{fluid} and \textit{residual} mass, and discuss the experimental results according to the ratio of these two quantities. We also discuss the pros and the cons of each behavior to the light of their energy saving potential, impact on scheduling and acceptability to the user. We hope that this work will pave the way for studies involving the user more intensely.

Future work will focus on (i)~improving the data center model to deal with the model simplifications listed in Subsection~\ref{sec:limits}, (ii)~proposing schedulers capable of leveraging the efforts made by the user (eg., through ``green SLA''), (iii)~elaborating on the user model to more realistically account for submission patterns and response to feedback from the infrastructure (as proposed by Feitelson~\cite{feitelsonResamplingFeedbackNew2016}) and (iv)~going beyond the limited scope of demand response to reason on the sustainability of the infrastructure as a whole.
%\jmp{une autre idée à garder en tête, pas à mettre là : étudier d'autres types de dégradation : x2 et /2 c'est bien, mais pourquoi pas x3 et /3 ou x4 et /4 ? Peut-être les résultats seraient différents. Aussi, dans le délai, on pourrait imaginer que l'user ne revienne pas à la fin de la time window mais le lendemain par exemple...}

\subsubsection{Acknowledgements} 
Experiments presented in this paper were carried out using the Grid'5000 testbed, supported by a scientific interest group hosted by Inria and including CNRS, RENATER and several Universities as well as other organizations (see \url{https://www.grid5000.fr}). This work was partly supported by the French Research Agency under the project Energumen (ANR-18-CE25-0008) and DataZero2 (ANR-19-CE25-0016).

%\jmp{j'ai trouvé bizarre de n'avoir aucune référence à des travaux de l'équipe. Dommage, non ?} 
%\geo{Peut être mettre un des articles de l'équipe où on fait 'juste' de l'opti énergétique de datacenter en tant que citation à l'endroit où ça parle d'effet rebond ? genre \url{https://oatao.univ-toulouse.fr/17206/1/villebonnet_17206.pdf}}
%\geo{Ou alors \url{https://hal.archives-ouvertes.fr/hal-02964970/document} où on fait des applications reconfigurables (qui ressemblent donc au modèle des tâches reconf ici), mais au niveau scheduler.}

\bibliographystyle{splncs04}
\bibliography{europar_2022}

\end{document}